\documentclass{article}
\usepackage{spconf,amsmath,graphicx}

\usepackage{amsfonts}
\usepackage{url}
\usepackage{multicol}
\usepackage{multirow}
\usepackage{hyperref}
\usepackage[font=small,skip=0pt]{caption}
\usepackage{subfig}
\usepackage{wrapfig}
\usepackage{cleveref}

\Crefname{section}{$\S$}{$\S$}

\title{Opening the Black Box of wav2vec Feature Encoder}

\name{
    Kwanghee Choi\textsuperscript{1}$^*$\thanks{$^*$Equal contributors.},
    Eun Jung Yeo\textsuperscript{2}$^*$
}

\address{
    Department of Computer Science and Engineering, Sogang University, Republic of Korea\textsuperscript{1} \\
    Department of Linguistics, Seoul National University, Republic of Korea\textsuperscript{2}
}

\begin{document}

\maketitle
\begin{abstract}
Self-supervised models, namely, wav2vec and its variants, have shown promising results in various downstream tasks in the speech domain.
However, their inner workings are poorly understood, calling for in-depth analyses on what the model learns.
In this paper, we concentrate on the convolutional feature encoder where its latent space is often speculated to represent discrete acoustic units.
To analyze the embedding space in a reductive manner, we feed the synthesized audio signals, which is the summation of simple sine waves.
Through extensive experiments, we conclude that various information is embedded inside the feature encoder representations: (1) fundamental frequency, (2) formants, and (3) amplitude, packed with (4) sufficient temporal detail.
Further, the information incorporated inside the latent representations is analogous to spectrograms but with a fundamental difference: latent representations construct a metric space so that closer representations imply acoustic similarity.
\end{abstract}


\section{Introduction} \label{sec:intro}
Large-scale self-supervised models, namely, wav2vec \cite{schneider2019wav2vec} and its variants \cite{baevski2020wav2vec,babu2021xls}, have succeeded in a wide range of downstream tasks, such as automatic speech recognition \cite{baevski2020wav2vec} and keyword spotting \cite{yang2021superb}.
Furthermore, recent advances demonstrate that the model can handle general tasks such as emotion recognition, environment classification, and music tagging, even with models pretrained with unlabeled speech data \cite{yang2021superb,wang2022towards}.
Concretely, self-supervised learning in wav2vec is designed as partially hiding the sequential representations and identifying the original representation based on its surroundings in a contrastive manner \cite{schneider2019wav2vec,baevski2020wav2vec}.
Hence, one cannot help but wonder about the acoustic capabilities of self-supervised learning, as it does not explicitly guide the model to capture the necessary information for speech analysis.

Though with subtle differences, wav2vec-like models usually consist of two components: the feature encoder based on consecutive 1D convolutional layers and the transformer-based contextual network \cite{schneider2019wav2vec}.
It is commonly understood as the convolutional feature encoder to embed acoustic features where the contextual network handles global, abstract information \cite{baevski2020wav2vec}.
Based on this understanding, state-of-the-art models choose between passing the raw waveform through the 1D convolutional feature encoder or directly inputting the spectrogram \cite{yang2021superb,wang2022towards}.
However, the inner workings are yet poorly understood, calling for in-depth analyses of what the convolutional net actually encodes.

In this paper, we are motivated to explore whether the convolutional feature encoder can substitute spectrograms, which is the basic tool in audio spectral analysis, showing frequency-wise amplitude information with sufficient temporal detail.
Hence, we start with the simplest form: the weighted sum of different sine waves.
By taking the reductionist approach, we synthesize inputs block by block instead of using the recorded speech signals. 
Our purpose is to provide a fundamental understanding of how and why the encoder works so well.

We summarize the rest of the paper as follows.
\Cref{sec:related} covers related works that analyze the inner workings of wav2vec-like models by leveraging existing datasets.
Then, we describe our novel analysis method in \Cref{ssec:settings}.
\Cref{ssec:temporal} explores the \textbf{temporal detail} encoded in the representations, focusing on whether the representations are consistent with different time steps.
\Cref{ssec:f0} further investigates the representation space with respect to \textbf{fundamental frequencies} (F0s) and biases, comparing with perceptual scales such as the Mel or the Bark scale.
\Cref{ssec:formants} dives into the delicacies of \textbf{formants} aided with linguistic analyses and compares different wav2vec variants.
Finally, \Cref{ssec:amplitude} covers how \textbf{amplitude} differences impact latent representations.
We conclude our paper with in-depth discussions on comparing spectrogram and feature encoder representations in \Cref{sec:discu}.

\section{Related Works} \label{sec:related}
In the past decades, using learnable convolutional layers to handle the raw audio signals has been thoroughly explored, where the prevalent idea was to mimic or enhance the spectrogram.
For instance, \cite{palaz2013estimating} observed frequency responses of a few convolutional filters, and concluded that the convolutional neural networks (CNNs) behave as filter matcher.
The follow-up paper showed that the early layers model the spectral envelope \cite{palaz2015analysis}.
In \cite{trigeorgis2016adieu}, the authors observed the activations directly, looking for correlations with the acoustic features such as energy or fundamental frequency.
Further analysis was conducted in \cite{hoshen2015speech}, which construct spectrogram-like visualizations based on activations and initialize CNN weights with filterbanks.
Similarly, \cite{thickstun2017learning} concentrated on the network architecture to mimic spectrograms.
Even though these works focused on bridging acoustic theory and neural nets, it was done before the emergence of wav2vec and self-supervised learning, urging the need for extended analysis.
Further, when previous studies often regarded CNN representations as the substitution for spectrograms, our analysis emphasizes the fundamental difference between the spectrogram and the self-supervised representations (\Cref{sec:discu}).

After the success of wav2vec-like self-supervised models, researchers concentrated on what high-level information transformers encode.
Linear probing, which attaches a shallow learnable layer to consume intermediate representations of a frozen transformer, was often used to verify whether the information of interest exists.
Accordingly, it was discovered that transformer layers from wav2vec-like models contain local acoustic, phonetic, word, language, and speaker information \cite{pasad2021layer,fan2021explore}.
Some also focused on self-supervised learning losses and verified that LSTMs can retain phoneme, speaker, gender, and language information with self-supervision \cite{niekerk2021analyzing,seyssel2022probing}.
In addition, \cite{choi2022temporal} observed that the transformer attentions focus on the region of interest even without direct supervision.
Though with promising analyses and results, studies have concentrated on the characteristics of the transformer, but not the convolutional feature encoder.
Moreover, studies are mostly limited to probing high-level, abstract information.

Others also focused on both convolutional feature encoders and contextual network.
For instance, \cite{vu2014investigating} compared the latent representation space of shallow neural networks with the vowel chart, showing the embedded formant and articulation information. 
\cite{tom2022wav2vec} conducted similar research, but analyzed the wav2vec model.
Concretely, the authors fine-tuned the pretrained model to the phone recognition task and observed both transformer and CNN representations.
Both retained the phonetic concepts, such as vowel space, but each contained different knowledge, CNN the gender information and transformer the language information.
However, there are two limitations on this research.
First, using recorded audio for analysis hinders understanding the representations in detail. 
Second, as the model is supervised with phoneme information, it is challenging to understand the effect of self-supervision separately.
Hence, this paper inputs synthesized audio to fully control the acoustic characteristics and analyze the pretrained model without any fine-tuning.

\section{Analysis} \label{sec:analysis}
\subsection{Experimental Settings} \label{ssec:settings}
\textbf{Models.}
We analyze wav2vec 2.0 base (\texttt{Base}, \cite{baevski2020wav2vec}), a self-supervised model trained with English speech.
However, to compare the differences between a bigger model or cross-lingual pretraining, we also compare wav2vec 2.0 large (\texttt{Large}, \cite{baevski2020wav2vec}) and XLS-R 300M (\texttt{XLS-R}, \cite{babu2021xls}), which has roughly the same size as \texttt{Large}.
\textbf{Architecture.}
For all three models, the feature encoder consists of five 1D convolution layers with layer normalization and activation functions.
The model is fed with 16kHz raw audio, and the effective window size and the stride for the feature encoder are 400 and 320, respectively.
For example, for 1-second input, the encoder outputs 49 (temporal) $\times$ 512 (feature) vector.
\textbf{Visualization.}
We use UMAP \cite{mcinnes2018umap} to visualize the embedding space.
As wav2vec 2.0 uses contrastive loss based on cosine similarity $S_c(*, *)$ \cite{mcinnes2018umap}, we choose the cosine distance $D_c = 1-S_c$ as the distance measure for UMAP.
Except for the distance measure, we follow the default settings.
\textbf{Implementation.}
We employ publicly released models and libraries of Huggingface \cite{wolf2020transformers}.
For more details, consult our repository.\footnote{\url{https://github.com/juice500ml/unbox-w2v-convnet}}

\subsection{Temporal detailedness} \label{ssec:temporal}
\textbf{Does the convolutional feature encoder yield consistent representations with different timesteps?}
We first verify whether the representations fluctuate with time when we feed a simple 1-second sine signal: $x(t) = \sin(2\pi f_0 t)$, where $f_0\in \{100, 200, \dots 500\textrm{Hz}\}$.
We denote the output representations as $\textbf{R} = [\textbf{r}_1 \textbf{r}_2 \dots \textbf{r}_T]$ where $T$ is the total timestep.

\begin{wrapfigure}{r}{0.2\textwidth}
  \centering
  \vspace{-1em}
  \includegraphics[width=0.22\textwidth]{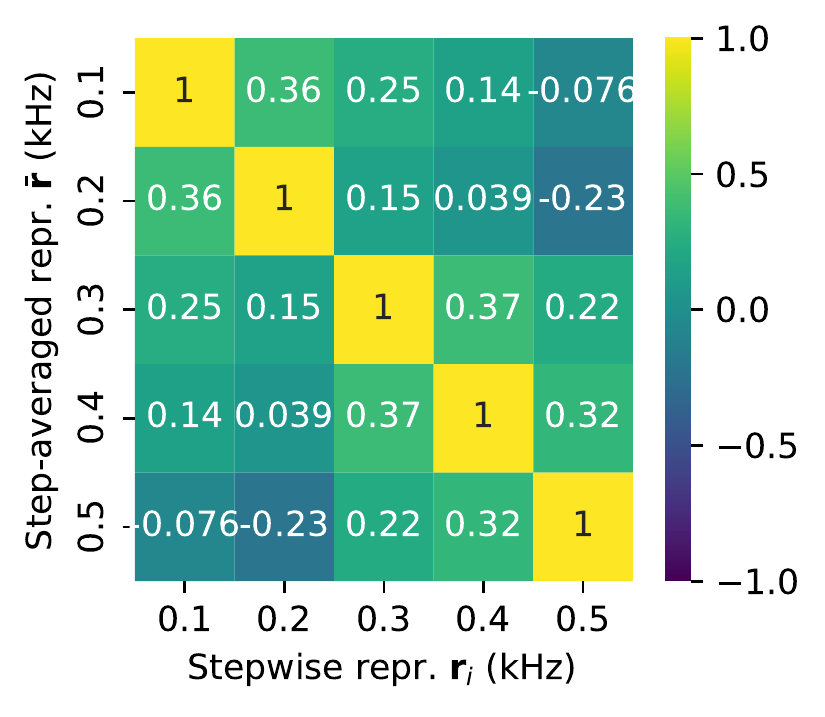}
  \vspace{-1em}
  \caption{Avg. cos similarity.}
  \vspace{-1em}
  \label{subfig:temporal-heatmap}
\end{wrapfigure}
We calculate the time-averaged representation $\Bar{\textbf{r}} = \tfrac{1}{T} \textstyle\sum_i \textbf{r}_i$ and the average cosine similarity between $\Bar{\textbf{r}}$ and $\textbf{r}_i$, $\tfrac{1}{T} \textstyle\sum_i S_c(\Bar{\textbf{r}}, \textbf{r}_i)$.
The row and the column of \cref{subfig:temporal-heatmap} denote averaged $\Bar{\textbf{r}}$ and stepwise representation $\textbf{r}_i$.
We observe that the average similarity between the averaged and the stepwise representations are $1.0$, the representations being consistent even as time varies.
The matrix is symmetric, which further strengthens the conclusion.
Hence, we use the averaged representations $\Bar{\textbf{r}}$ in the remaining experiments.

\textbf{How much temporal detail does the encoder handle?} \begin{wrapfigure}{r}{0.25\textwidth}
  \centering
  \vspace{-1em}
  \includegraphics[width=0.25\textwidth]{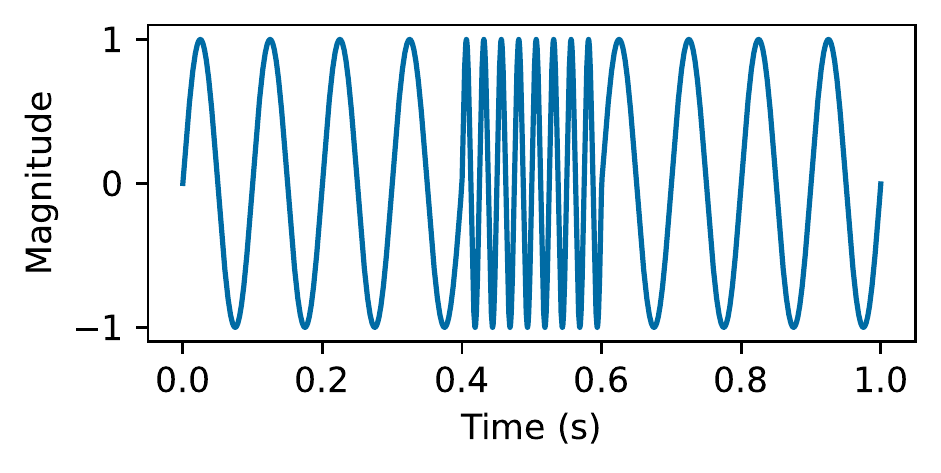}
  \vspace{-1em}
  \caption{Mixed signal example.}
  \vspace{-1em}
  \label{subfig:temporal-example}
\end{wrapfigure}
To observe whether the encoder is sensitive to the change in signals, we mix two sine signals (\cref{subfig:temporal-example}).
We put a high frequency signal with $f_0$=800Hz in the middle of the 1-second signal of lower $f_0$=200Hz.
We vary the duration of the high frequency signal from 320ms to 10ms and compare the stepwise representations $\mathbf{r}_i$ with the averaged representations $\bar{\mathbf{r}}$ separately obtained from clean 1-second signal.

\begin{figure}[t] 
\centering
\includegraphics[width=0.98\columnwidth]{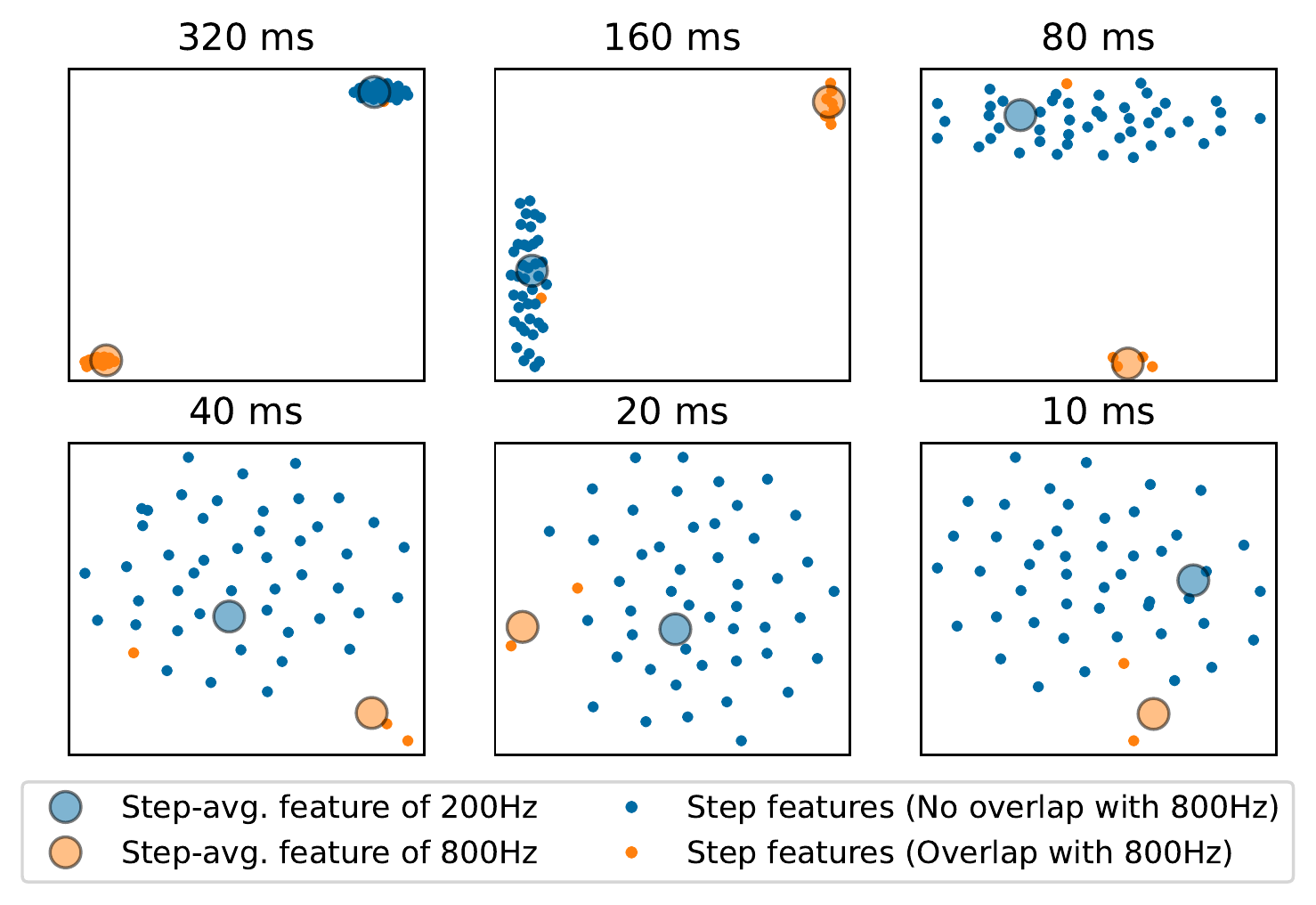}
\caption{
Stepwise representations of the mixed signal.
Considering the window and the stride size of the encoder, we color-code based on whether the step signal overlaps with the window of the specific time step.
Note that some steps may overlap more than others.
}
\label{subfig:temporal-umap}
\vspace{-1.5em}
\end{figure}

We observe in \cref{subfig:temporal-umap} that the stepwise representation differs when $f_0$ changes, even for short bursts.
However, its level of distinction decreases in shorter periods.
As the popular setting in speech processing for feature extraction is a window size and stride of 25ms and 10ms \cite{rabiner2007introduction}, we can conclude that the level of detailedness is adequate for speech analysis.

\vspace{-0.5em}\subsection{Fundamental frequency} \label{ssec:f0}
\textbf{How is the fundamental frequency embedded in the representation space?}
To explore how different fundamental frequency $f_0$ impacts representations, we use the 1-second signal $x(t) = A_0 \sin(2\pi f_0 t)$ with $A_0 = 1.0$ and $f_0$ ranging from 10Hz to 8kHz with 10Hz interval.
As the input signal for the wav2vec 2.0 has a sample rate of 16kHz, we set the maximum as the Nyquist sampling rate of 8kHz.

\begin{figure}[t] 
\centering
\subfloat[Representation space with\\varying $f_0$.]{\includegraphics[width=0.49\columnwidth]{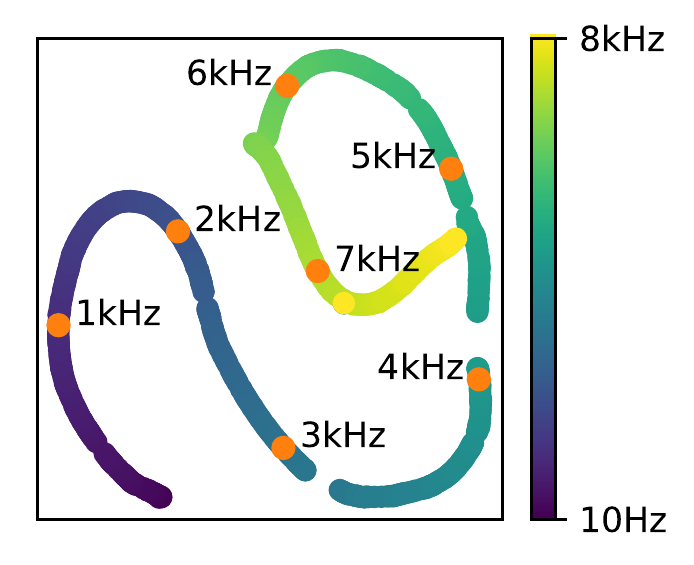}\label{subfig:f0-repr}}
\subfloat[Comparing different frequency scales.]{\includegraphics[width=0.49\columnwidth]{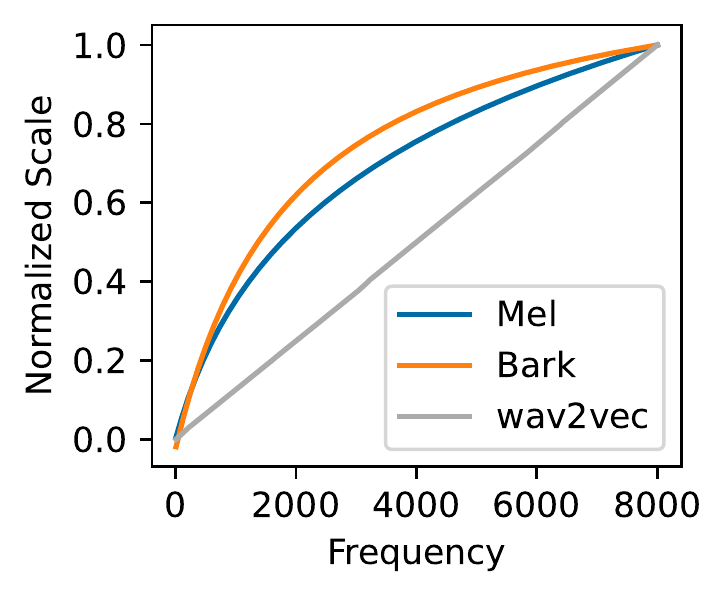}\label{subfig:f0-scale}}
\caption{
Representation analyses on varying fundamental frequency.
}
\label{figs:f0}
\vspace{-1em}
\end{figure}

\Cref{subfig:f0-repr} demonstrates that not only do signals with similar F0 merely cluster with each other but are linearly ordered.
This result aligns with the previous findings that the speech representation of the encoder is clustered by gender \cite{seyssel2022probing,tom2022wav2vec}, where the same gender often has a similar F0 range.

We further analyze the cosine distance of the neighboring F0s by comparing it with existing frequency scales, Mel and Bark, in \cref{subfig:f0-scale}.
We accumulate the cosine distance of the neighboring signals to construct a scale, as representations farther apart is analogous to using a larger filter bank.
We observe that the scale is evenly spaced, unlike the perception-inspired human-made scales.
Hence we conclude that the wav2vec 2.0 embeds F0 not only in an orderly fashion but in a linear way.
Moreover, the embeddings are well-put in the full range of frequencies, supporting the effectiveness of the wav2vec 2.0 on various auditory downstream tasks \cite{yang2021superb,wang2022towards}.

\textbf{Does the encoder ignore the signal bias?}
Similar to the previous setting (\cref{subfig:temporal-heatmap}), we feed the encoder with the following 1-second signal: $x(t) = A_0 \sin(2\pi f_0 t) + b$, where we set $A_0 = 0.5$ with varying bias $b \in (-0.5, 0.25, 0.0, 0.25, 0.5)$.

\begin{wrapfigure}{r}{0.2\textwidth}
  \centering
  \vspace{-0.5em}
  \includegraphics[width=0.2\textwidth]{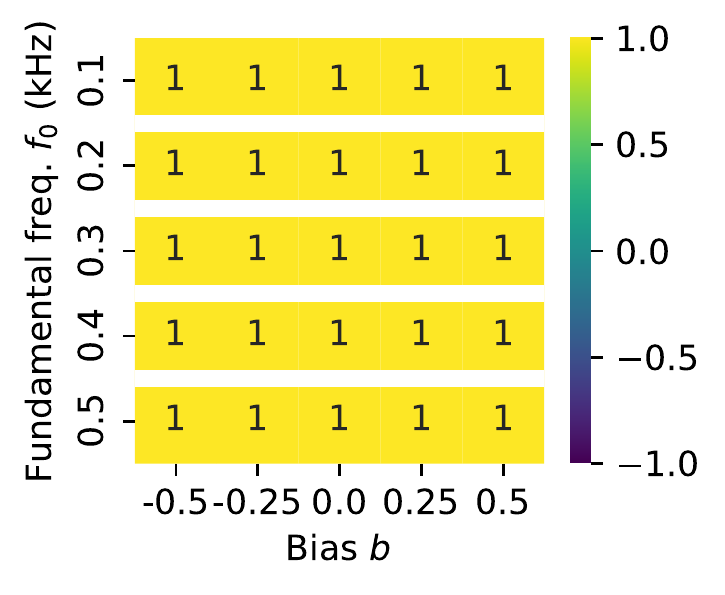}
  \vspace{-1em}
  \caption{Cos. similarity with varying $f_0$ and $b$.}
  \vspace{-2em}
  \label{subfig:f0-bias}
\end{wrapfigure}
We compare the cosine similarity between time-averaged features with and without bias ($b_0 = 0.0$) for multiple $f_0$s in \cref{subfig:f0-bias}.
We see no difference across all $f_0$s.
We suspect layer normalization inside the encoder enables this trait.
\vspace{-2em}

\subsection{Formants: F1 and F2} \label{ssec:formants}
To observe the impact of formants, we feed the signal $x(t)=A_0 \sin(2\pi f_0 t) + A_1 \sin(2\pi f_1 t) + A_2 \sin(2\pi f_2 t)$ to the encoder, where $A_0=0.5, A_1=0.35, A_2=0.15$.
We set $f_0 \in \{100, 125, \dots, 225\mathrm{Hz}\}$, and check $30$ evenly spaced frequencies for both $f_1 \in [235\mathrm{Hz}, 850\mathrm{Hz}]$ and $f_2 \in [595\mathrm{Hz}, 2400\mathrm{Hz}]$, based on the lowest and highest F1 and F2 formants of \cite{catford1988practical}.

\begin{wrapfigure}{r}{0.2\textwidth}
  \centering
  \vspace{-1em}
  \includegraphics[width=0.2\textwidth]{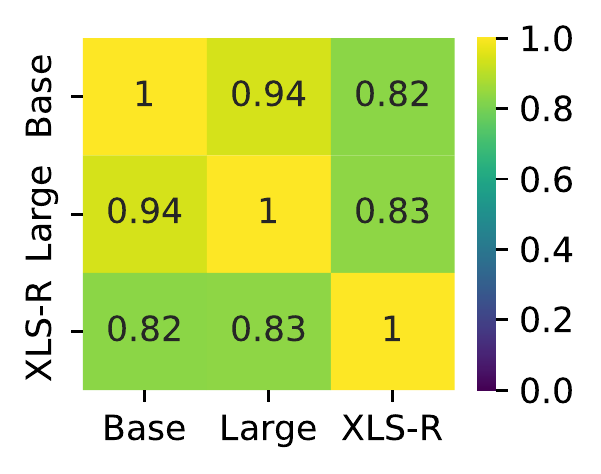}
  \vspace{-1em}
  \caption{Difference between representations.}
  \vspace{-1em}
  \label{subfig:formant-cka}
\end{wrapfigure}
\textbf{Comparing different encoders.}
We quantify the differences by feeding the signals defined above to the three different encoders. We measure the distances by the much used linear Centered Kernel Alignment (CKA) \cite{kornblith2019similarity} in \cref{subfig:formant-cka}.
All encoders are similar, whereas the difference based on the dataset is much more significant than that of model size.
Hence we omit the \texttt{Base} model plots due to space constraints, but we qualitatively confirmed the similarity between the \texttt{Base} and \texttt{Large} representations.

\begin{figure}[t] 
\centering
\subfloat{\includegraphics[width=0.24\textwidth]{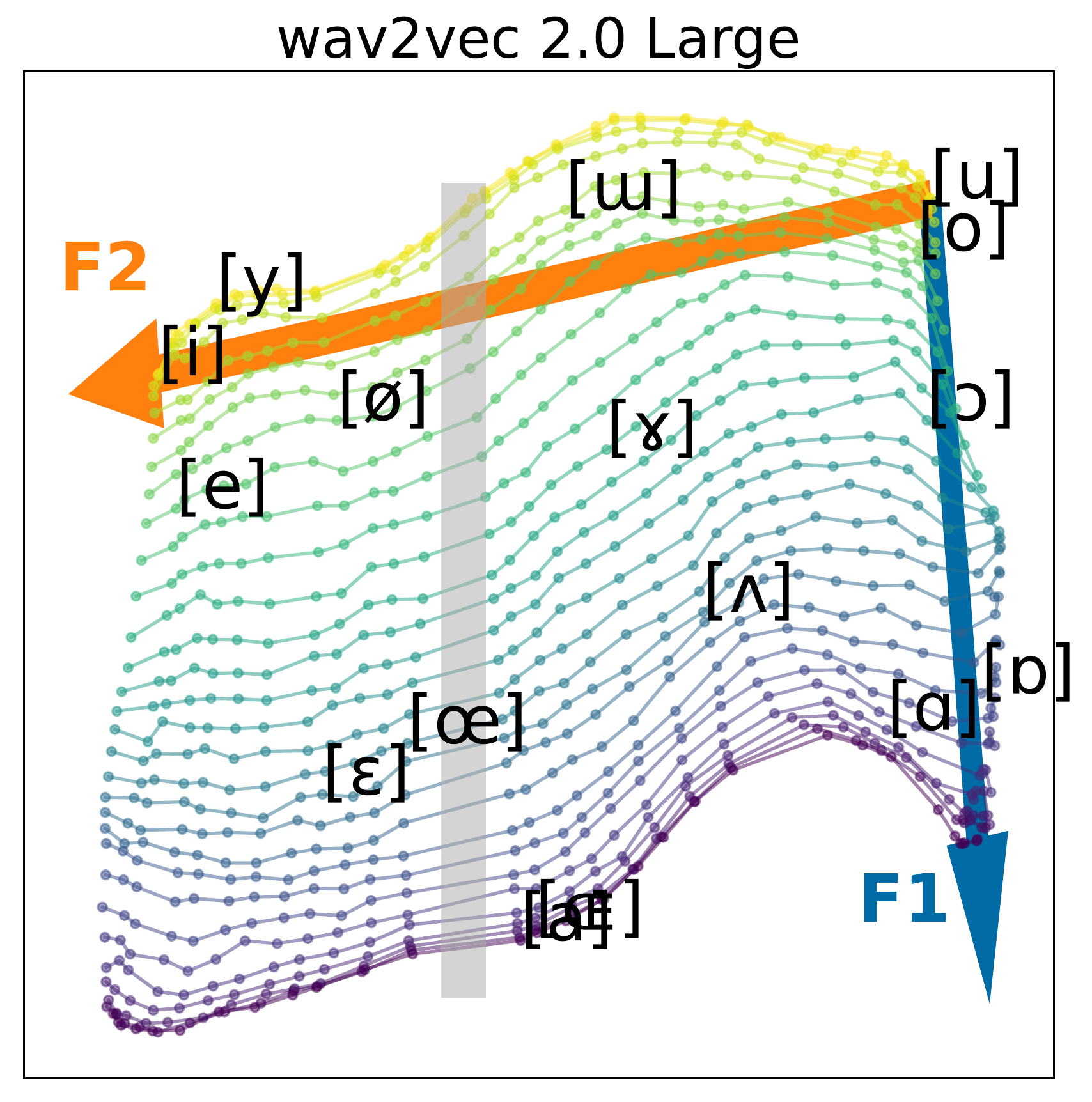}}
\subfloat{\includegraphics[width=0.24\textwidth]{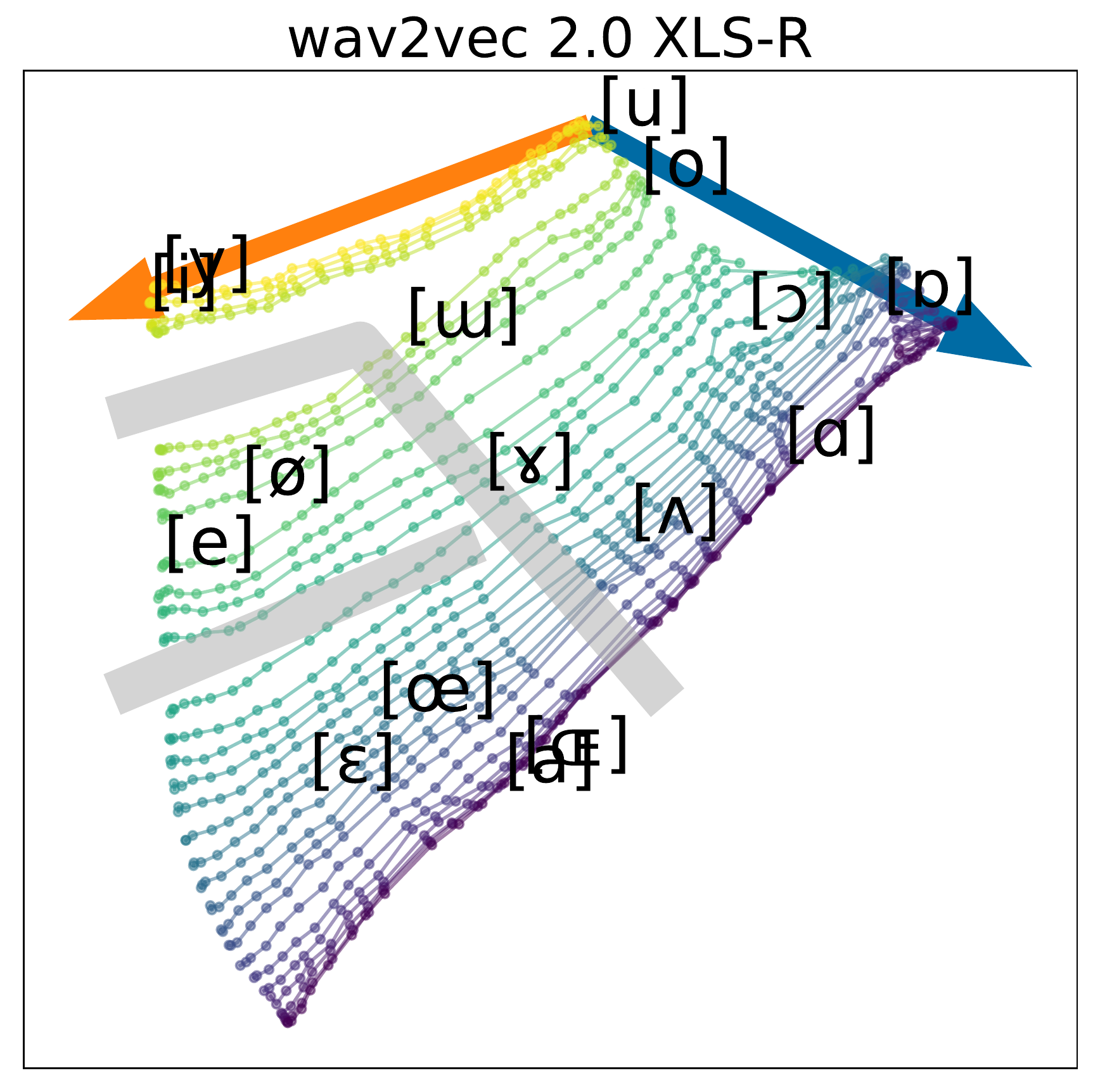}}
\caption{
Representation space of the two models on varying F1 and F2 formants.
Same color denotes same F1.
Blue (F1) and orange (F2) arrows indicate increasing frequencies.
Cardinal vowels with their standard F1 and F2 values \cite{catford1988practical} are plotted.
We color the distinctively sparse regions gray.
}
\label{fig:formant-f1f2}
\vspace{-2em}
\end{figure}

\textbf{How is vowel space constructed?}
We first fix $f_0=120\textrm{Hz}$ to concentrate on the impact of F1 and F2.
\Cref{fig:formant-f1f2} presents the F1-F2 representation space for \texttt{Large} and \texttt{XLS-R}.
To intuitively understand how encoders embed formant information, we compare the representation space of the two models.
We can observe that the representations construct a grid structure; not only are they clustering, but the manifold is smoothly formed, reducing the burden of the transformer-side.
Furthermore, we found three linguistic findings.
First, the space for /œ/ is empty in \texttt{Large} model while \texttt{XLS-R} model successfully handles the formants for the corresponding phoneme. 
This can be explained by different phoneme inventories: while English does not contain /œ/ speech sound, XSL-R is pre-trained with languages which has /œ/ phoneme, such as French.
Second, representations close to corner vowels /i/, /u/, and /a/ are strongly clustered for \texttt{XLS-R}, compared to \texttt{Large}.
This reflects the characteristics of the corner vowels, which represent the boundaries of the vowel space, meaning no vowels exist outside these vowels.
Lastly, discriminations between different vowels stand out for \texttt{XLS-R} compared to \texttt{Large}.
Concretely, one huge gap is observed between front vowels and others for \texttt{Large}, while \texttt{XLS-R} representations also embeds differences of vowels' height into high, mid, and low.
We attribute this phenomenon to \texttt{XLS-R} hearing a wider range of phonemes in various languages.

\begin{figure}[t] 
\centering
\subfloat{\includegraphics[width=0.24\textwidth]{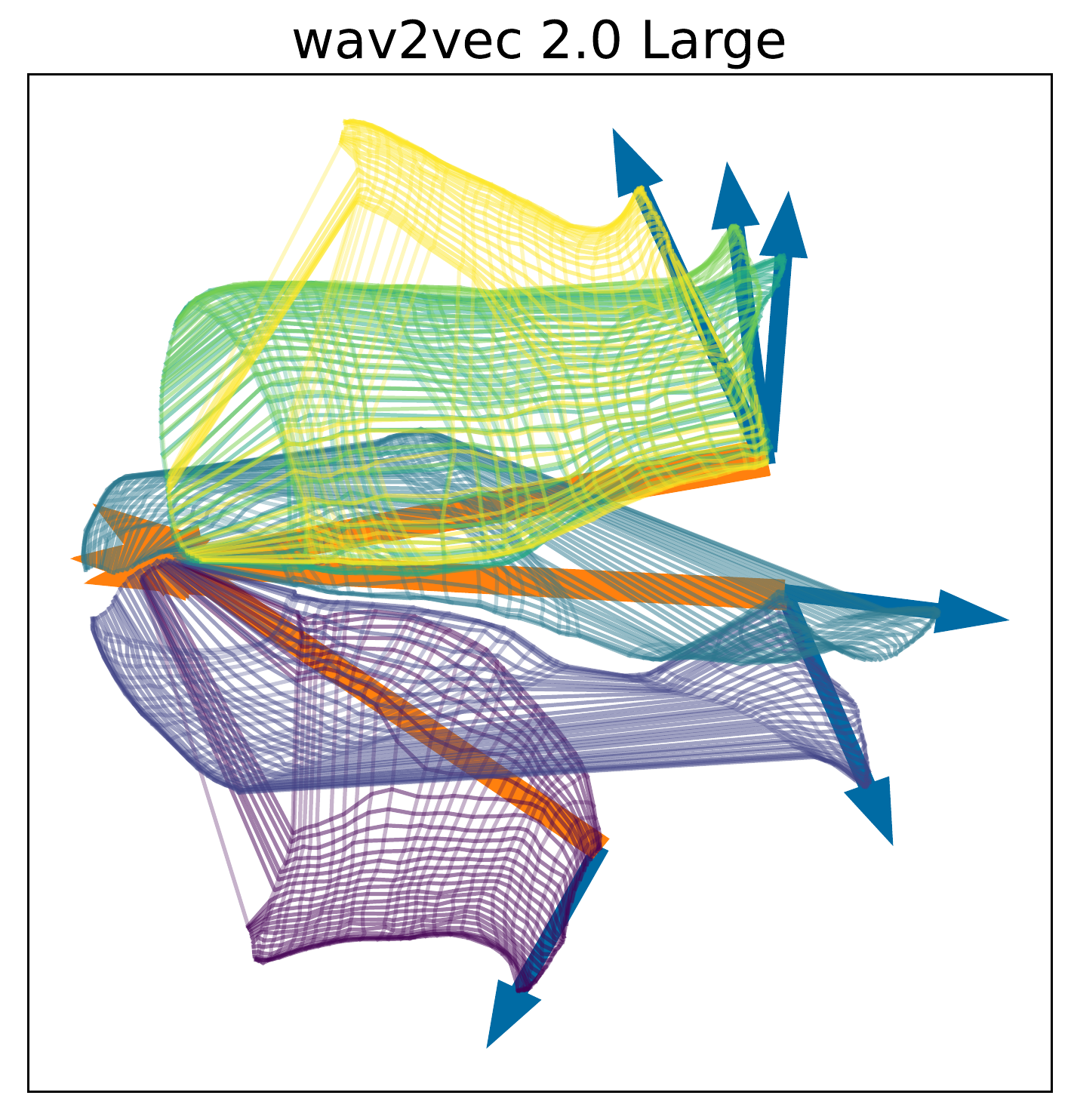}}
\subfloat{\includegraphics[width=0.24\textwidth]{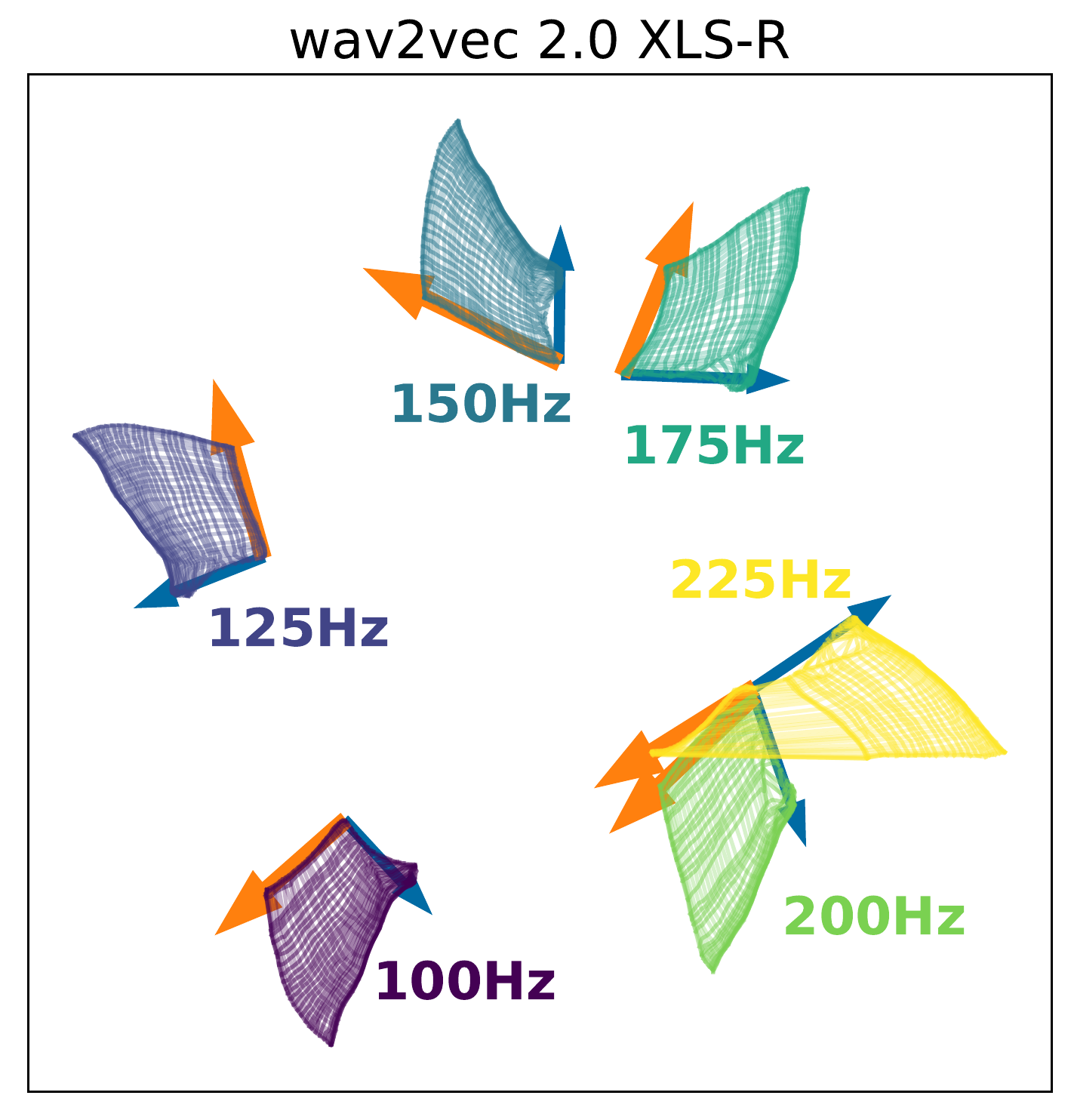}}
\caption{
Representation space of the two models on varying F0, F1, and F2.
Same color denotes same F0.
Blue (F1) and orange (F2) arrows indicate increasing frequencies.
}
\label{fig:formant-f0f1f2}
\vspace{-2em}
\end{figure}

\textbf{F0 vs Formants.}
We extend the previous experiment to show how F0, F1, and F2 interact within the representation space.
We can observe even though F1 and F2 still maintain the grid structure, F0 separates each grid. 
This implies that F0's influence is bigger than that of F1 and F2.
However, unlike \texttt{XLS-R} which scatters different F0 signals, \texttt{Large} tends to group F0s into high and low, which seems to indicate gender difference.
We suspect that this is due to the pretraining dataset for \texttt{Large} being more speech-oriented, whereas \texttt{XLS-R} uses various datasets that partially contain non-speech.

\begin{figure}[t] 
\centering
\subfloat{\includegraphics[width=0.23\textwidth]{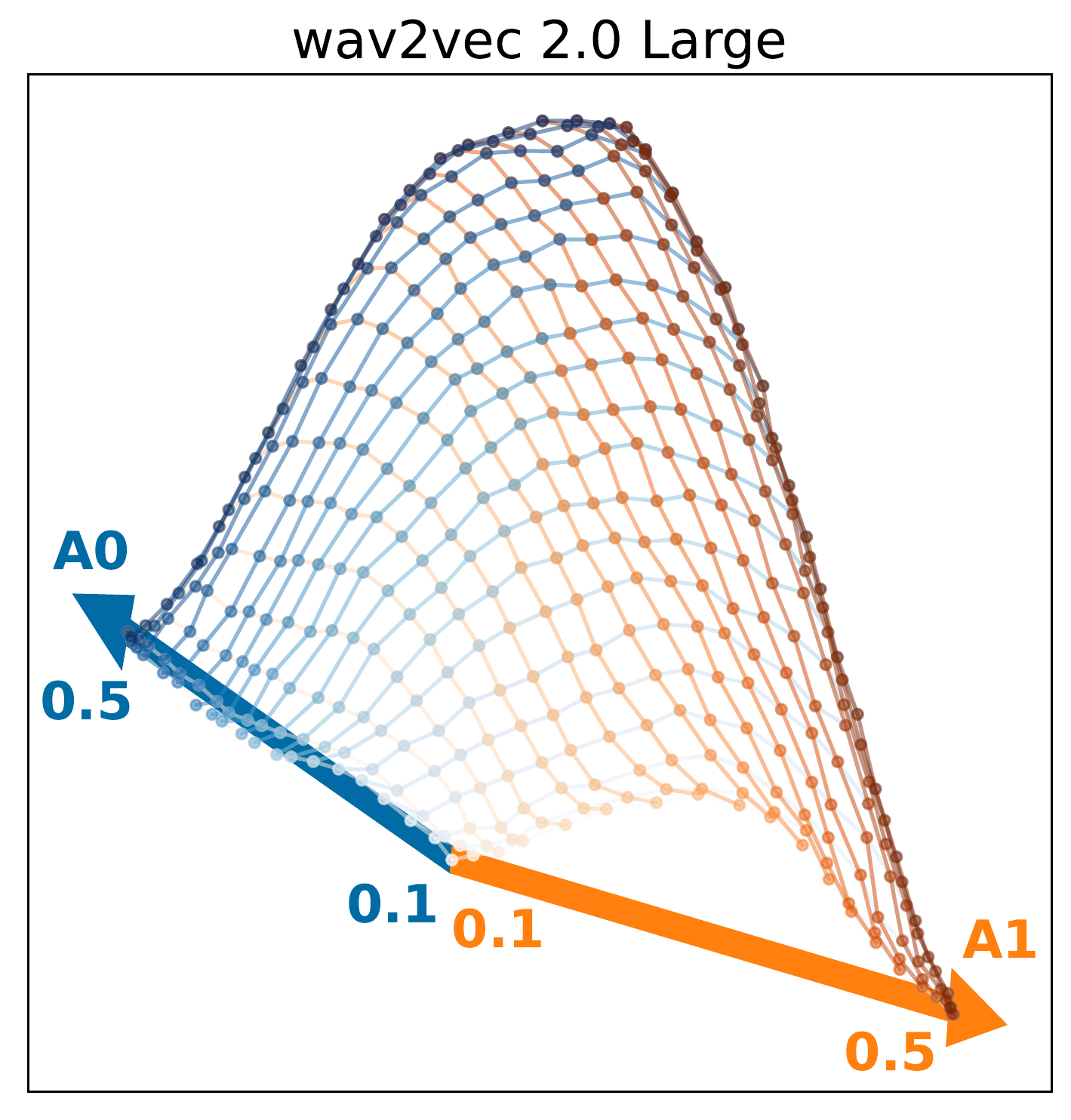}}
\subfloat{\includegraphics[width=0.23\textwidth]{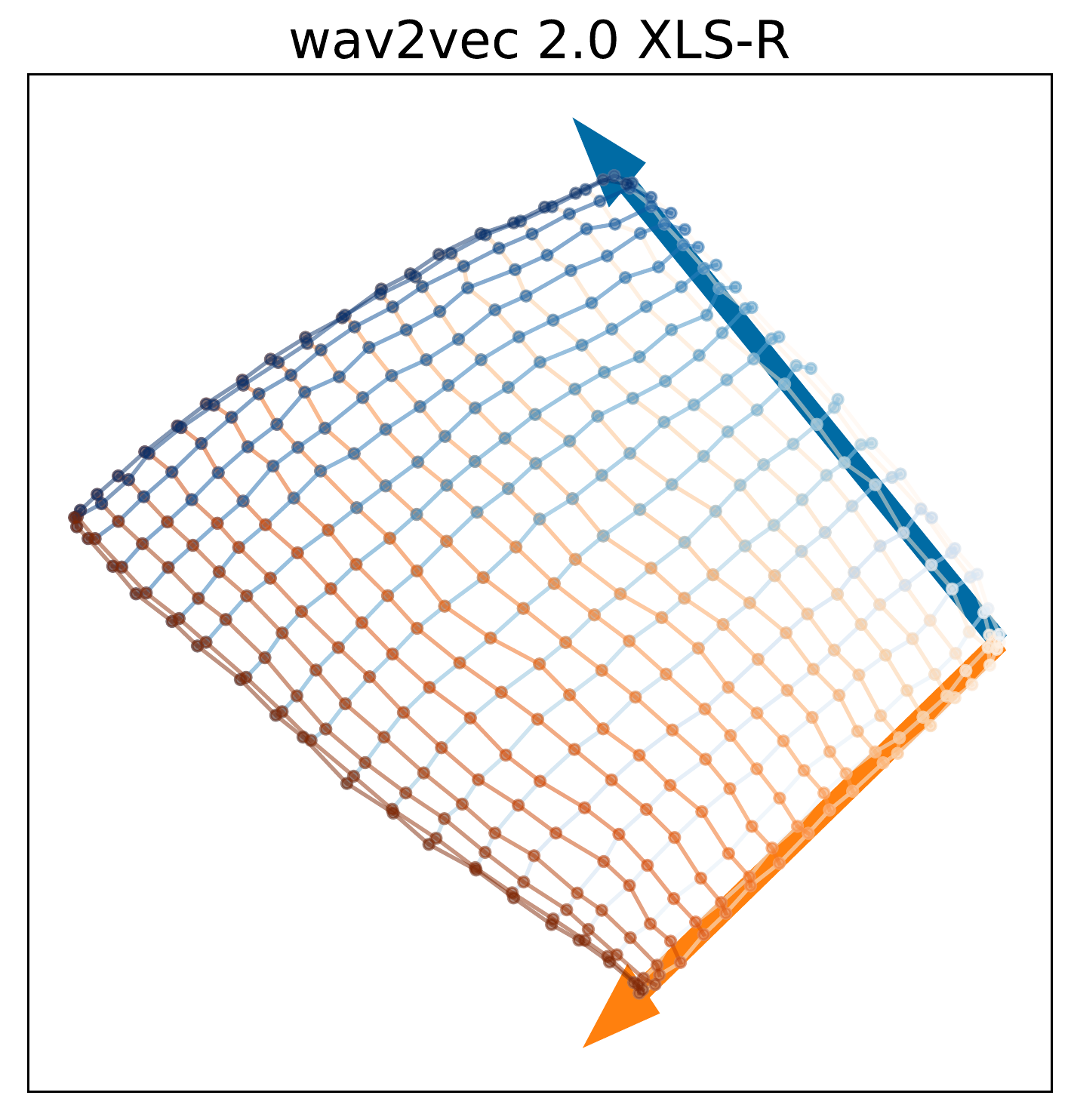}}
\caption{
Representation space of the two models on varying amplitudes.
We color-code based on the amplitudes $A_0$ (Blue) and $A_1$ (Orange), where it gets darker as the amplitude gets bigger.
}
\label{fig:amplitude}
\vspace{-1em}
\end{figure}

\vspace{-0.5em} \subsection{Amplitude} \label{ssec:amplitude} \vspace{-0.5em}
\textbf{How does the encoder handle when signals with different amplitudes are added?}
Similar to \Cref{ssec:formants}, we use the 1-second signal: $x(t) = A_0\sin(2\pi f_0 t) + A_1\sin(2\pi f_1 t)$, where $f_0$=100Hz, $f_1$=700Hz, and varying $A_0$ and $A_1$ from $0.1$ to $0.5$.
We evenly space out regarding energy by making the interval equal for squared amplitude.

Similar to F1-F2 (\cref{fig:formant-f1f2}), we verify that the amplitude is also stored in a grid structure in \cref{fig:amplitude}.
However, unlike \texttt{XLS-R}, \texttt{Large} tends to cluster the low-energy signals.
Similar to \Cref{ssec:formants}, we suspect this due to the data distribution; noisy sources would contain more diverse signal energies.

\vspace{-0.5em} \section{Discussion and Conclusion} \label{sec:discu} \vspace{-0.5em}
Throughout various analyses, we demonstrate that all the information inside the spectrogram is successfully embedded in the representation space of the convolutional encoder.
The representations are temporally consistent and sufficiently detailed (\Cref{ssec:temporal}), fundamental frequencies lie in an orderly and linear way (\Cref{ssec:f0}), bias is ignored (\Cref{ssec:f0}), formants are stored in a grid-like structure, linguistically interpretable way (\Cref{ssec:formants}), and amplitudes are also kept grid-like (\Cref{ssec:amplitude}).

However, we emphasize that our work does not aim to show the representational equivalence between the feature encoder and the spectrogram.
The former store information by \textbf{relative distances} from others, whereas the latter encodes with absolute values.
For example, consider $x_i(t)=\sin(2\pi f_it)$ where $f_0$=100Hz, $f_1$=200Hz, and $f_2$=8kHz.
The distances of spectrogram-induced features are the same for $D(x_1, x_2)$ and $D(x_1, x_3)$, but highly different for the feature encoder.
Hence, we conclude that encoder representations construct an \textbf{acoustic metric space}, similar to word representations like word2vec \cite{mikolov2013linguistic}.
This revelation aligns with self-supervision; contrastive and diversity losses being based on the relative distance between representations.
Further, transformer's attention mechanisms also depend on the relative distance; hence using the encoder will be more advantageous.
A natural progression of this work is to design theory-driven sound representations, improving upon the current data-driven methods.

\bibliographystyle{IEEEbib}
\bibliography{refs}

\end{document}